\documentclass[conference]{IEEEtran}
\usepackage{amsmath,amssymb,amsfonts}
\usepackage{algorithmic}
\usepackage{hyperref}
\usepackage{textcomp}
\usepackage{xcolor}
\usepackage{graphicx}
\usepackage{dblfloatfix}
\usepackage{rotating}
\usepackage{savesym}
\savesymbol{checkmark}
\usepackage{dingbat}
\usepackage{booktabs}
\usepackage{cite}

\usepackage[ruled]{algorithm2e} 
\makeatletter
\newcommand{\removelatexerror}{\let\@latex@error\@gobble}
\makeatother
\def\tightlist{}

\begin{document}

\title{Specimens as research objects: reconciliation across distributed repositories to enable metadata propagation.}

\author{\IEEEauthorblockN{1\textsuperscript{st} Nicky Nicolson}
\IEEEauthorblockA{\textit{Royal Botanic Gardens, Kew} \\
and \textit{Brunel University, London}\\
London, UK \\
n.nicolson@kew.org}
\and
\IEEEauthorblockN{2\textsuperscript{nd} Alan Paton}
\IEEEauthorblockA{\textit{Royal Botanic Gardens, Kew}\\
London, UK \\
a.paton@kew.org}
\and
\IEEEauthorblockN{3\textsuperscript{rd} Sarah Phillips}
\IEEEauthorblockA{\textit{Royal Botanic Gardens, Kew}\\
London, UK \\
sarah.phillips@kew.org}
\and
\IEEEauthorblockN{4\textsuperscript{th} Allan Tucker}
\IEEEauthorblockA{\textit{Brunel University, London}\\
London, UK \\
allan.tucker@brunel.ac.uk}
}

\maketitle

\begin{abstract}
Botanical specimens are shared as long-term consultable research objects in a
global network of specimen repositories. Multiple specimens are generated from a
shared field collection event; generated specimens are then managed individually
in separate repositories and independently augmented with research and
management metadata which could be propagated to their duplicate peers.
Establishing a data-derived network for metadata propagation will enable the
reconciliation of closely related specimens which are currently dispersed,
unconnected and managed independently. Following a data mining exercise applied
to an aggregated dataset of 19,827,998 specimen records
from 292 separate specimen repositories,
36\% or 7,102,710 specimens 
are assessed to participate in duplication relationships, allowing the propagation 
of metadata among the participants in these relationships, totalling:
93,044 type citations,
1,121,865 georeferences,
1,097,168 images and 2,191,179
scientific name determinations. The results enable the creation of networks to
identify which repositories could work in collaboration. Some classes of
annotation (particularly those regarding scientific name determinations)
represent units of scientific work: appropriate management of this data would
allow the accumulation of scholarly credit to individual researchers: potential
further work in this area is discussed.
\end{abstract}


\section{Introduction}

Botanical specimens are core research objects in the science of taxonomy
(the naming of biological organisms), stored for long term consultation
in institutional repositories and referenced in academic works.
Worldwide there are 3,001 herbaria (botanical specimen repositories),
containing 387,007,790 specimens - representing collections gathered
over hundreds of years \cite{thiers_worlds_2018}. Due to their
physical characteristics (flattened, dried plant material is typically
mounted on a large sheet of paper, stored inside a paper folder) and
their management as a long term, consultable record, specimens act as
vehicles for the communication of results and theories, as researchers
annotate the paper sheet underlying the specimen. Annotations placed on
specimen sheets are public and available for use by other researchers,
this public yet potentially unpublished status is discussed in
\cite{conn_information_2003}.

Taxonomic researchers populate institutional repositories by conducting
collection events (usually field-based) which generate multiple
specimens. Recommended botanical practice is for a single collection
event to generate five to six specimens, which will be deliberately
distributed to separate institutional repositories. Physical
distribution of specimens has three main goals: to maximise access -
researchers working on their local flora should be able to consult the
relevant specimens in their national herbarium, to provide resilient
storage - duplicate specimens insure against disastrous loss of a single
repository, and to ensure efficient use of storage space within
repositories \cite{bridson_herbarium_1998}. Duplicate specimens are
also used in genetic analyses: if the samples were collected from
separate individuals, the duplicate set can be used to assess genetic
diversity across the sampled population. Scientific theories regarding
the recognition of species and their interrelationships are developed by
researchers as they work with the specimens, which are traditionally
accessed either by loan or by visits to institutions; more recently
specimen digitisation initiatives have enabled online access to specimen
metadata records and high quality images, this simplifies search and
retrieval of specimens and associated metadata, and allows some level of
specimen examination to be conducted remotely. Independent creation and
management of metadata for specimen duplicates can be inefficient
(metadata creation is repeated unnecessarily), and inadvertently
misleading (metadata diverges between different members of a specimen
duplicate group).

One particular class of research annotation is the application of a
scientific name to the specimen: this may be an existing name, or the
researcher may recognise that the specimen represents a new species.
Species description in plants is ongoing with circa two thousand new
plant species described each year
\cite{noauthor_international_nodate}. When a new species is
described, one specimen is chosen as a physical representation of the
otherwise abstract scientific name. Specimens which formally represent a
scientific name are called type specimens; the selection of these is
called type citation. When a specimen is cited as a type, all peers
(``duplicates'') which are generated from the same collection event -
but which may be stored and managed remotely, in separate repositories -
are also considered to have type status. New scientific names are
created via a formal publication process governed by the International
Code of Nomenclature for algae, fungi and plants
\cite{mcneil_international_2012}. The majority of new species are
discovered from historic specimens already lodged within specimen
repositories, being formally described years after collection
\cite{bebber_herbaria_2010}. The use of duplicate specimens as
vehicles for the communication of results is illustrated by the historic
use of ``exsiccatae''. These are uniform specimen sets with information
displayed on printed labels distributed to multiple herbaria, and until
1953 were considered a valid publication mechanism for new scientific
names \cite{triebel_online_2011}
\cite{mcneil_international_2012}.

Taxonomists consider type specimens to be the most valuable specimens in
a repository, and management reporting often includes both the total
number of specimens held and the number of type specimens. The first
major digitisation effort in botany (JSTOR Global Plants Initiative)
focussed on the digitisation of type specimens across more than 300
institutions in over 70 countries \cite{ithaka_jstor_2015}. In
addition to reporting on the total numbers of specimens and types housed
in an institutional repository \cite{bras_french_2017}, managers are
also interested in the numbers of new type citations published each year
as a metric of on-going research use of their specimens
\cite{friis_how_2012}. Some natural history institutions have
experimented with bibliometrics to quantify use of their specimens in a
publication context \cite{winker_natural_2013}.

In addition to their core use in the science of taxonomy, specimens
provide physical ``what, where, when'' evidence and are used for a wide
range of scientific applications such as species distribution modelling
\cite{chapman_uses_2005}. Specimen exchange networks have also been
used for historical social network analysis
\cite{groom_herbarium_2014}. These applications are generally
dependent on aggregations of specimen metadata mapped to a common data
standard and sourced from many different institutional repositories.

\textbf{Problem statement} Despite the widespread recognition that
botanical specimens form a global collection, there is currently no flow
of data from the point of creation (via the field collection event) to
the generated specimens wherever they may be located for long term
storage. Despite advances in the mobilisation and standardized
representation of specimen metadata across the different specimen
repositories, duplicate specimens have so far gone undetected, with
metadata records for duplicates appearing unlinked in aggregated
datasets. The main data elements needed to assess specimens as
potentially arising from a shared collection event - collector name,
along with the collector's recordnumber and eventdate - are not formally
managed. These missing links mean that valuable research annotations and
type citations are not easily shared between repositories, and impacts
all downstream users of specimen data: taxonomic researchers working
with individual specimens are unable to benefit from knowledge added
elsewhere, leading to misinterpretation due to inaccurate and/or out of
date naming, and users working with large aggregations of specimen data
can find that specimen number estimates are overstated, as their
datasets contain hidden duplicates.

The research described in this paper applies machine learning to a set
of aggregated specimen metadata to identify and reconcile the collectors
responsible for the creation of specimens, enabling the detection and
linkage of specimen duplicates generated from the field work of the
identified collectors. In contrast to existing work on annotation
propagation - which has focussed on potential changes in working
practices and tools and techniques to enable and incentivize this
\cite{suhrbier_annosysimplementation_2017}
\cite{macklin2006developing} - this work applies these techniques to
a dataset of existing digitally available specimen data in order to
calculate the numbers of existing metadata elements and annotations
which may be propagated between separate institutional repositories.

The remainder of this paper is structured as follows: a background
section further introduces the problem domain with an explanation of the
specimen life cycle and the kinds of annotations applied at each stage,
and worked examples of distributed specimen sets whose members are
independently managed at different institutions. Materials and methods
describes the application of a machine learning process to a dataset of
specimen data from the Global Biodiversity Information Facility to
identify specimen duplicates. Criteria for the identification and
assessment of duplicate sets are proposed. The resulting specimen
duplicate analysis is used to answer the following questions:

\begin{enumerate}
\def\labelenumi{\arabic{enumi}.}
\tightlist
\item
  How many distributed, independently managed specimens can be
  reconciled across separate institutional repositories and linked as
  generated products of a common collection event?
\item
  How many metadata elements and research annotations can be propagated
  between institutional specimen repositories?
\item
  Can specimen duplicate linkages be used to infer network relationships
  between institutional repositories, which institutions are most
  frequently linked and do sub-communities or cliques exist in the
  inferred network?
\end{enumerate}

Preliminary results are presented and ideas for expansion and future
work are proposed.

\section{Background}

This section outlines the stages in the specimen life cycle, and
indicates relevant projects at each stage.

\textbf{Collection and storage}: these activities represent standard
practice across the specimen repositories

\begin{itemize}
\tightlist
\item
  \textbf{Collection}: material is gathered from the field and details
  of the collection locality (associated species, geology, habitat etc)
  are recorded in the collectors field notebook. The collectors
  recordnumber provides the cross-reference between the data recorded in
  the field notebook and the physical material collected, this is
  usually a sequential number managed individually by the collector.
\item
  \textbf{Accessioning}: material is received by a specimen repository
  and prepared for long term storage, including mounting on a sheet of
  paper (for dried specimens).
\end{itemize}

\textbf{Digitisation}: due to the number of specimens held in the global
collection, digitisation is incomplete, and is progressing through a
variety of cross-cutting institutional, regional, international and
thematic projects. The JSTOR Global Plants Initiative selected a
particular class of specimens for digitisation (type specimens) across
300 institutions \cite{ithaka_jstor_2015}, other projects have been
set up to digitise all specimens gathered from a particular country to
enable data repatriation, as in the Brazilian REFLORA programme
\cite{noauthor_reflora_nodate} and to digitise specimens held within
a particular country as in the US National Science Foundation funded
Advancing Digitisation of Biocollections programme
\cite{noauthor_advancing_nodate}. These latter projects show a trend
of government funding for digitisation, recognising that these are part
of the national scientific infrastructure \cite{bras_french_2017}
\cite{page_digitization_2015}.

\begin{itemize}
\tightlist
\item
  \textbf{Databasing}: details of the specimen (metadata) are added to
  an institutional data repository.
\item
  \textbf{Aggregation}: databased records can be mapped to a data
  standard (e.g.~Darwin Core \cite{wieczorek_darwin_2012}) and
  shared with aggregation projects. The \href{www.gbif.org}{Global
  Biodiversity Information Facility} is an intergovernmental
  organisation that aggregates specimen-derived species occurrence
  records (alongside records from observations) to facilitate scientific
  research, \href{www.idigbio.org}{iDigBio} is a US based aggregator
  which focusses only on specimen derived data.
\item
  \textbf{Georeferencing}: the metadata record in the institutional
  repository can have latitude and longitude added (this may be a costly
  step for historic records where the original collection locality is
  only a textual description of the place). Economies of scale are
  possible if records can be ordered so that similar places are
  georeferenced together \cite{hill_location_2009}
  \cite{garcia2010data}.
\item
  \textbf{Imaging}: the specimen is imaged and a reference to the image
  is added to the metadata. If the specimen metadata is shared with an
  aggregator the digital image may also be mobilised.
\end{itemize}

Depending on their range of holdings, some institutions are involved in
multiple digitisation projects, others not at all. With technical
advances in digitisation and the setup of high-throughput imaging
facilities, some of these steps may be performed out of sequence - if
the digitisation project is of a sufficient scale, it may be cost
effective to rapidly image the specimens first and perform the metadata
capture later, from a high quality digital image
\cite{van_oever_pilot_2012} \cite{heerlien_natural_2015}
\cite{sweeney_large-scale_2018}.

\textbf{Use as a research object}: these steps outline the use of the
specimen as a taxonomic research object. The use of specimens as a data
source for computational applications such as species modelling is
covered in the digitisation steps above, digitisation steps also
facilitate discovery and access of specimens for taxonomic research.
Annotation mobilisation work has focussed on tooling for the collection
and propagation of newly generated annotations, including the projects
AnnoSys \cite{suhrbier_annosysimplementation_2017} and Filtered Push
\cite{macklin2006developing}. There has also been an effort to
standardise the citation of specimens so that different repositories use
a common HTTP URI based naming convention by which their digital
metadata records can be accessed \cite{guntsch_actionable_2017}. By
convention, the citation of specimen records irrespective of
digitisation status is made by stating the collector name, number and
date, along with the herbarium code \cite{thiers_worlds_2018} in
which the physical specimen may be found. These kinds of references can
be found throughout the botanical literature, and examples are shown in
the worked examples in the next section.

\begin{itemize}
\tightlist
\item
  \textbf{Determination}: the specimen is labelled with a scientific
  name, the date and the name of the researcher who made the
  determination are also added.
\item
  \textbf{Citation}: the specimen is cited in a published academic work
  (e.g.~to evidence the presence of a species in a geographic region).
\item
  \textbf{Type citation}: the specimen is referenced as a type specimen
  in a published academic work to create a new species name.
\end{itemize}

The long term creation of a global network of specimen repositories, the
more recent efforts to enable virtual access to specimens and their
metadata, and the practice of sharing research annotations all fit well
with the FAIR principles for scientific data management
\cite{wilkinson_fair_2016}: ensuring that the metadata and specimens
on which scientific analyses are based are Findable, Accessible,
Interoperable and Retrievable.

\subsection{Worked examples}

\begin{table*}[ht!]
    \centering
    \caption{Worked examples}\label{tab:worked_examples}
    \resizebox{\textwidth}{!}{%
\begin{tabular}{llllllllll}
\toprule
 recordedBy                          & recordNumber   & eventDate   & scientificName                & institutionCode   & referenced in publication   & digitised   & typestatus   & georeferenced   & imaged     \\
\midrule
 P. F. Zika                          & 26185          & 2013-06-09  & Sedum citrinum Zika           & BH                & \checkmark                  & -           & -            & -               & -          \\
 Zika, Peter F.                      & 26185          & 2013-06-09  & Sedum citrinum Zika           & CAS               & \checkmark                  & \checkmark  & \checkmark   & \checkmark      & -          \\
 Peter F. Zika                       & 26185          & 2013-06-09  & Sedum citrinum Zika           & CAS-BOT-BC        & -                           & \checkmark  & -            & -               & -          \\
 P. F. Zika                          & 26185          & 2013-06-09  & Sedum citrinum Zika           & CHSC              & -                           & \checkmark  & -            & \checkmark      & -          \\
 P. F. Zika                          & 26185          & 2013-06-09  & Sedum citrinum Zika           & GH                & \checkmark                  & -           & -            & -               & -          \\
 Zika, P.F.                          & 26185          & 2013-06-09  & Sedum citrinum Zika           & K                 & -                           & \checkmark  & \checkmark   & -               & \checkmark \\
 P. F. Zika                          & 26185          & 2013-06-09  & Sedum citrinum Zika           & MO                & \checkmark                  & -           & -            & -               & -          \\
 P. F. Zika                          & 26185          & 2013-06-09  & Sedum citrinum Zika           & NY                & -                           & \checkmark  & \checkmark   & \checkmark      & \checkmark \\
 P. F. Zika                          & 26185          & 2013-06-09  & Sedum citrinum Zika           & OSC               & \checkmark                  & -           & -            & -               & -          \\
 Peter F. Zika                       & 26185          & 2013-06-09  & Sedum citrinum Zika           & RSA               & \checkmark                  & \checkmark  & -            & \checkmark      & -          \\
 Peter F. Zika                       & 26185          & 2013-06-09  & Sedum citrinum Zika           & UC                & \checkmark                  & \checkmark  & -            & \checkmark      & -          \\
 P. F. Zika                          & 26185          & 2013-06-09  & Sedum citrinum Zika           & US                & \checkmark                  & \checkmark  & \checkmark   & -               & \checkmark \\
 P. F. Zika                          & 26185          & 2013-06-09  & Sedum citrinum Zika           & WTU               & \checkmark                  & -           & -            & -               & -          \\
 \midrule
 P. C. Hutchison \& J. K. Wright      & 5738           & 1964-06-19  & Solanum sanchez-vegae S.Knapp & F                 & \checkmark                  & \checkmark  & \checkmark   & -               & \checkmark \\
 P. C. Hutchison \& J. K. Wright      & 5738           & 1964-06-19  & Solanum aligerum Schltdl.     & F                 & -                           & \checkmark  & -            & -               & -          \\
 Hutchison, P.C.                     & 5738           & 1964-06-19  & Solanum sanchez-vegae S.Knapp & K                 & \checkmark                  & \checkmark  & \checkmark   & \checkmark      & \checkmark \\
 Paul C. Hutchison|J. Kenneth Wright & Hutchison 5738 & 1964-06-19  & Solanum cutervanum Zahlbr.    & MO                & -                           & \checkmark  & -            & -               & -          \\
 P. C. Hutchison                     & 5738           & 1964-06-19  & Solanum sanchez-vegae S.Knapp & NY                & -                           & \checkmark  & \checkmark   & \checkmark      & \checkmark \\
 P. C. Hutchison                     & 5738           & 1964-06-19  & Solanum sanchez-vegae S.Knapp & NY                & -                           & \checkmark  & \checkmark   & \checkmark      & \checkmark \\
 P.C. Hutchison \& J.K. Wright        & 5738           & 1964-06-19  & Solanum sanchez-vegae S.Knapp & P                 & \checkmark                  & -           & -            & -               & -          \\
 P. C. Hutchison \& J. K. Wright      & 5738           & 1964-06-19  & Solanum sanchez-vegae S.Knapp & US                & \checkmark                  & \checkmark  & \checkmark   & -               & \checkmark \\
 P.C. Hutchison \& J.K. Wright        & 5738           & 1964-06-19  & Solanum sanchez-vegae S.Knapp & USM               & \checkmark                  & -           & -            & -               & -          \\
\bottomrule
\end{tabular}
    }
\end{table*}

This section is intended to illustrate the problem statement - that
specimen duplicates are (1) widely present in distributed specimen
repositories, (2) unidentified in data aggregations built by combining
specimen datasets and (3) that specimen metadata attached to derived
specimens generated from a single source can diverge due to separate and
independent data curation practices. Two examples have been selected,
representing the two extremes of species description citing botanical
specimens: species discovery in-field formalised by rapid publication
just one year after collection, and species discovery in-repository with
formalised description decades after field collection. A considerable
proportion of new species are described from material already collected
and stored in specimen repositories \cite{bebber_herbaria_2010}. The
second example shows a species description occurring 46 years after the
field collection of the plant material on which is it based.

For each example we will assemble a dataset of potential specimens,
which is constructed as the superset of the specimens referenced in the
literature (which may or may not be digitised) and the relevant specimen
records found in digital form in a data aggregator. We then examine the
metadata attached to the specimens, showing where this has diverged due
to independent management. These are shown in table
\ref{tab:worked_examples}.

\subsubsection{Example 1: Rapid publication of species discovered
in-field}

See table \ref{tab:worked_examples}, example 1. (Table data source:
\href{http://api.gbif.org/v1/occurrence/search?recordnumber=26185\&eventDate=2013-06-09}{gbif.org})

The publication data (displayed below) shows that there are at least 9
specimen duplicates, stored in different institutional repositories,
indicated by the capitalised alphabetic herbarium codes (WTU, BH etc
\cite{thiers_worlds_2018}). The exclamation mark (!) after a code is
a convention to indicate that the author has actually seen the specimen.
In this case the author is also the collector of the specimen, so all
are listed as having been seen.

\begin{quote}
\textbf{\emph{Sedum citrinum}} Zika, \emph{sp. nov.}
\textbf{Type}:---UNITED STATES. California: Del Norte County, ridge 1.4
air km north of South Red Mountain, 1050 m, 9 June 2013, \emph{P. F.
Zika 26185} (holotype: WTU!; isotypes: BH!, CAS!, GH!, MO!, OSC!, RSA!,
UC!, US!). \cite{zika_new_2014}
\end{quote}

There are 8 digitally available records for this set of specimens, drawn
from 8 separate institutional specimen repositories. These are
independently managed and not interlinked. Despite being generated from
the same collection event, the specimen metadata show variation due to
isolated management in separate repositories: 5 of the 8 are
georeferenced, 4 of the 8 specify a type status and 3 of the 8 have an
associated image. We can therefore calculate that the group contains
propagable annotations for georeferences, typestatus and image
(i.e.~that for each annotation class, the group contains records with
and without the annotation set, meaning that the annotation could be
propagated from the specimens with the annotation to their peers without
it). Of the digitised specimens in the group: 3 could receive a
georeference, 4 could receive a type status annotation and 5 could be
linked to an associated image. The creation of a specimen group could
also make the initial creation of the specimen records for the currently
undigitised members more efficient, by using existing data as a starting
point rather than independently re-creating it.

\subsubsection{Example 2: Species discovery in-repository}

See table \ref{tab:worked_examples}, example 2. (Table data source:
\href{http://api.gbif.org/v1/occurrence/search?recordnumber=5738\&eventDate=1964-06-19}{gbif.org})

The publication data (displayed below) shows that there are at least 6
specimen duplicates, stored in 5 different institutional repositories.
The author has supplied a numeric identifier for some of the specimens
(shown in square brackets), to help the reader locate the relevant
records in specimen repository and / or its associated metadata
catalogue(s).

\begin{quote}
\textbf{\emph{Solanum sanchez-vegae}} S.Knapp, \emph{sp. nov.}
{[}urn:lsid:ipni.org:names:77103635-1{]} \textbf{Type}: Peru. Amazonas:
Prov. Chachapoyas, W side of Cerros Calla-Calla, 45 km above Balsas,
mid-way on road to Leimebamba, 3100 m, 19 Jun 1964, \emph{P.C. Hutchison
\& J.K. Wright 5738} (holotype, USM; isotypes, F {[}F-163831{]}, K
{[}K000545365{]}, P {[}P00549320{]}, US {[}US-246605{]}, USM).
\cite{knapp_four_2010}
\end{quote}

There are 7 digitally available records for this set of specimens, from
5 separate institutional specimen repositories. These are independently
managed and not interlinked. As per the first example, despite being
generated from the same collection event, the specimen metadata show
variation due to isolated management in separate repositories, with all
annotation categories holding inconsistent information: 3 of the 7 are
georeferenced, 5 of the 7 specify a type status, 5 of the 7 have an
associated image and 2 of the 7 have an outdated scientific name. We can
therefore calculate that of the 7 digitised specimens in the group: 4
could receive a georeference, 2 could receive a type status annotation
and 2 could be linked to an associated image.

These two different examples both show that the separate specimen
records held in different specimen repositories hold divergent metadata,
and that there is the potential for metadata propagation between members
of a specimen group. Specimen groups can be identified by grouping on
the collector, their field-assigned record number and the eventdate, but
this is non-trivial due to the variation in the recording style of the
collecting team (shown in the recordedBy column), as duplicate records
have been independently digitised to different data standards in
different institutions and projects.

\section{Materials and methods}

\subsection{Data}

A dataset of specimen data relating to vascular plants (those with
specialised tissues for the transport of water, encompassing ferns and
allied groups, and all seed plants) was downloaded from GBIF
\cite{gbif_trach_spec_2018} in Darwin Core
\cite{wieczorek_darwin_2012} archive format. This was input into a
data mining process based on the clustering technique DBSCAN in order to
detect collector entities \cite{nicolson_identifying_2017}. Specimen
records are eligible for data mining if they have a numeric component in
their \emph{recordnumber} (the sequential number managed by an
individual collector and assigned to field collection events), a precise
date recorded to the level of day (\emph{eventdate}), and a collector
name (\emph{recordedby}). The data mining process augments the specimen
dataset with a numeric identifier for the primary collector of the
specimen represented in the metadata record. This allows data to be
grouped as the product of the work of a particular collector,
irrespective of the lexical variation in the transcription of the
collectors names.

\begin{procedure*}
	\caption{detectDuplicateGroups(Specimens)}\label{proc:detect_dupl_groups}
	\KwIn{Specimens}
	\KwOut{LabelledSpecimens}
    	let \textit{$S$} be Specimens, the set of specimens to be grouped \\
        let \textit{$DuplicateGroups$} be S grouped by s.collector\_id, s.eventdate, s.recordnumber \\
        Apply an identifier to each group \\
        \For{$i\leftarrow 1$ \KwTo $|DG|$}{
            $dg \leftarrow DG[i]$ \\
            \For{$s$ in $dg$}{
            	$s.duplicate\_group\_id \leftarrow i$ \\
                LabelledSpecimens.append(s)
            }
        }
\end{procedure*}

\begin{procedure*}
	\caption{assessDuplicateGroups(LabelledSpecimens)}\label{proc:assess_dupl_groups}
	\KwIn{LabelledSpecimens}
	\KwOut{AssessedLabelledSpecimens}
    let \textit{$DuplicateGroups$} be LabelledSpecimens grouped by duplicate\_group\_id \\
     \For{$dg$ in $DuplicateGroups$}{
     	\For{$assessment\_field$ in $\{countrycode,order,family\}$}{
        	Create a new boolean field [assessement]\_conservative, which is set to \textit{True} \\
            if all members of the duplicate group share a single value for this field \\
        	$assessment\_values \leftarrow []$ \\
            \For{$s$ in $dg$}{
            	assessment\_values.append(s[assessment\_field])
            }
            $dg[assessment\_conservative] \leftarrow |assessment\_values| == 1$ \\
            Copy the assessment flag down to specimen level \\
            \For{$s$ in $dg$}{
            	$s.assessment\_conservative \leftarrow dg.assessment\_conservative$ \\
                AssessedLabelledSpecimens.append(s)
            }
      	}
    }
\end{procedure*}

\begin{procedure*}
	\caption{findPropagableAnnotations(assessedLabelledSpecimens)}\label{proc:count_annots}
	\KwIn{AssessedLabelledSpecimens}
	\KwOut{AssessedLabelledCountedSpecimens}
    let \textit{$DuplicateGroups$} be AssessedLabelledSpecimens grouped by duplicate\_group\_id \\
     \For{$dg$ in $DuplicateGroups$}{
     	let \textit{s} be the set of specimens included in \textit{dg} \\
     	Annotation fields are Boolean flags indicating if the specimen has this annotation set \\
     	\For{$annotation\_field$ in $\{georef, typestatus, image\}$}{
            $dg[annotation\_propagable] \leftarrow any(s.annotation\_field)\ and\ not\ all(s.annotation\_field)$\\
            Copy the propagable flag down to specimen level \\
            \For{$s$ in $dg$}{
            	$s.annotation\_propagable \leftarrow dg.annotation\_propagable$ \\
                AssessedLabelledCountedSpecimens.append(s)
            }
        }
     }

\end{procedure*}

\subsection{Detection of duplicate groups}

A group of specimens are asserted to be generated from a single
collection event if they share the same collector identifier (the
results of the collector data mining exercise), eventdate (when the
field collection event was carried out) and collector-assigned record
number. The record number has any alphabetic prefixes stripped from the
value - this normalises values which are sometimes presented with the
surname of the collector in the recordnumber field (see the worked
example in \ref{tab:worked_examples}).

See procedure listing \ref{proc:detect_dupl_groups}. The input into this
algorithm is a tabular data structure where each row represents a
specimen, with fields for collector\_id, eventdate and recordnumber.

\subsection{Establishing a confidence measure}

A confidence measure is applied to candidate duplicate groups by
examining the range of variation in fields within the duplicate group.
Three assessments are made, a spatial assessment using the countrycode
field (duplicate specimen records originating from the same collection
event should logically be located in the same country) and two taxonomic
assessments using the order and family fields. Biological taxonomy uses
a hierarchical system, where species are arranged into families, and
families into orders. Although a specimen may be re-determined (have
different scientific names applied to it) during its lifetime in a
specimen repository, it is less likely to be re-determined across higher
taxonomic boundaries. These flags detect variation in these higher-level
categories within a duplicate group.

Three Boolean flags were created (one for each assessment field), these
were set to True if all members of the candidate duplicate group share
the same value of the assessment field. All possible combinations of
these three flags were used to assess the duplicate groups. Only
duplicate groups meeting the most conservative assessment criteria
(where all of the assessment flags are True, indicating no variation in
these fields within the duplicate group) were carried forward for use in
subsequent analyses.

See procedure listing \ref{proc:assess_dupl_groups}. The input into this
algorithm is a tabular data structure where each row represents a
specimen, with fields for duplicate\_group\_id, countrycode, order and
family. This is the labelled output from the preceding algorithm
\ref{proc:detect_dupl_groups}.

\subsection{Assessing annotation status per specimen and detecting
groups with uneven annotation statuses}

Boolean flags were created to indicate if the specimen is georeferenced,
if the specimen has an associated image, and if the specimen is recorded
as having type status. \emph{Typestatus} values were used as described
in \cite{bebber_big_2012}.

For each annotation examined, two new Boolean fields were created on the
aggregated dataset - these are set to True if \emph{all} specimens in
the duplicate group have the annotation set and if \emph{any} specimens
in the duplicate group have the annotation set. A group is said to have
propagable annotations if it has any and not all annotations set for the
specimens within the group. Two count fields were also created for each
annotation, these were set to hold the number of specimens within the
group with and without the annotation set. The number of specimens which
could receive propagable annotations was determined by totalling the
number of specimens within groups with propagable annotations which did
not themselves have the annotation set.

See procedure listing \ref{proc:count_annots}. The input into this
algorithm is a tabular data structure where each row represents a
specimen, with a field for duplicate\_group\_id and a set of Boolean
fields to indicate the presence of annotations on the specimen
(georeference, typestatus, image). This is the assessed, labelled output
from the preceding algorithm \ref{proc:assess_dupl_groups}.

\subsection{Repository relationship analysis}

The sharing of specimens in a duplicate group implies a relationship
between the two (or more) institutional repositories participating in
the group. In this analysis, the data are reshaped to build a graph data
structure where nodes are institutional repositories and links are
created between a pair of nodes if the corresponding repositories share
specimens in a duplicate group. The links are weighted by the number of
groups shared. The resulting data structure is a weighted, undirected
graph. This inferred network data structure is visualised in Gephi
\cite{ICWSM09154}, using an OpenOrd \cite{martin2011openord}
layout following modularity analysis \cite{blondel_fast_2008} for
community detection.

\section{Results}

\subsection{Data mining}

The initial dataset downloaded from GBIF contained 63,492,620 records,
of these 19,827,998 records were eligible to be input into the data
mining process to detect the collector. The data mining process resulted
in 19,489,798 specimen records being labelled with an identifier for the
collector.

\subsection{Duplicate identification and assessment}

\begin{figure}
\caption{Assessment flag combination counts}\label{fig:assessment_graph}
\centering
\includegraphics[width=\columnwidth]{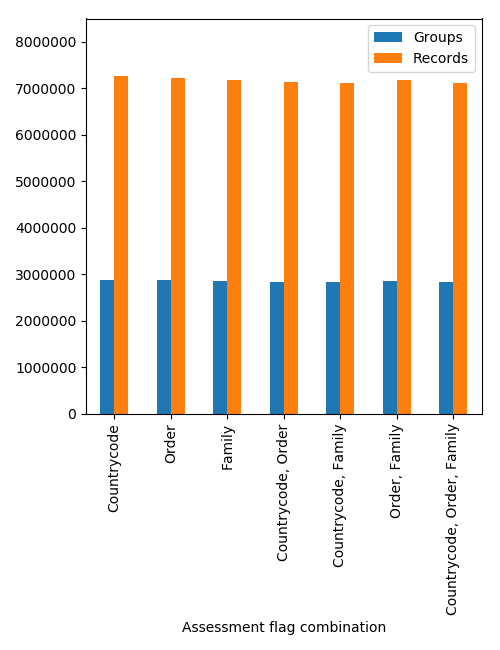}
\end{figure}

Of the 19,489,798 data mined records, 7,347,705 records participate in a
duplicate relationship, forming 2,914,181 duplicate groups. All
combinations of assessment flags with associated group and record counts
are depicted in figure \ref{fig:assessment_graph}.

\begin{figure}
\caption{Duplicate group sizes}\label{fig:dupl_group_sizes}
\centering
\includegraphics[width=\columnwidth]{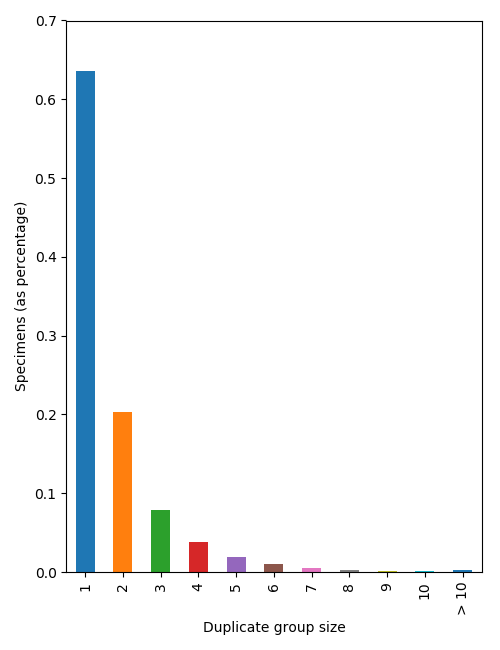}
\end{figure}

\begin{figure*}
\centering
\includegraphics[clip, trim={0.5cm 2.5cm 0.5cm 2.5cm}, scale=0.9]{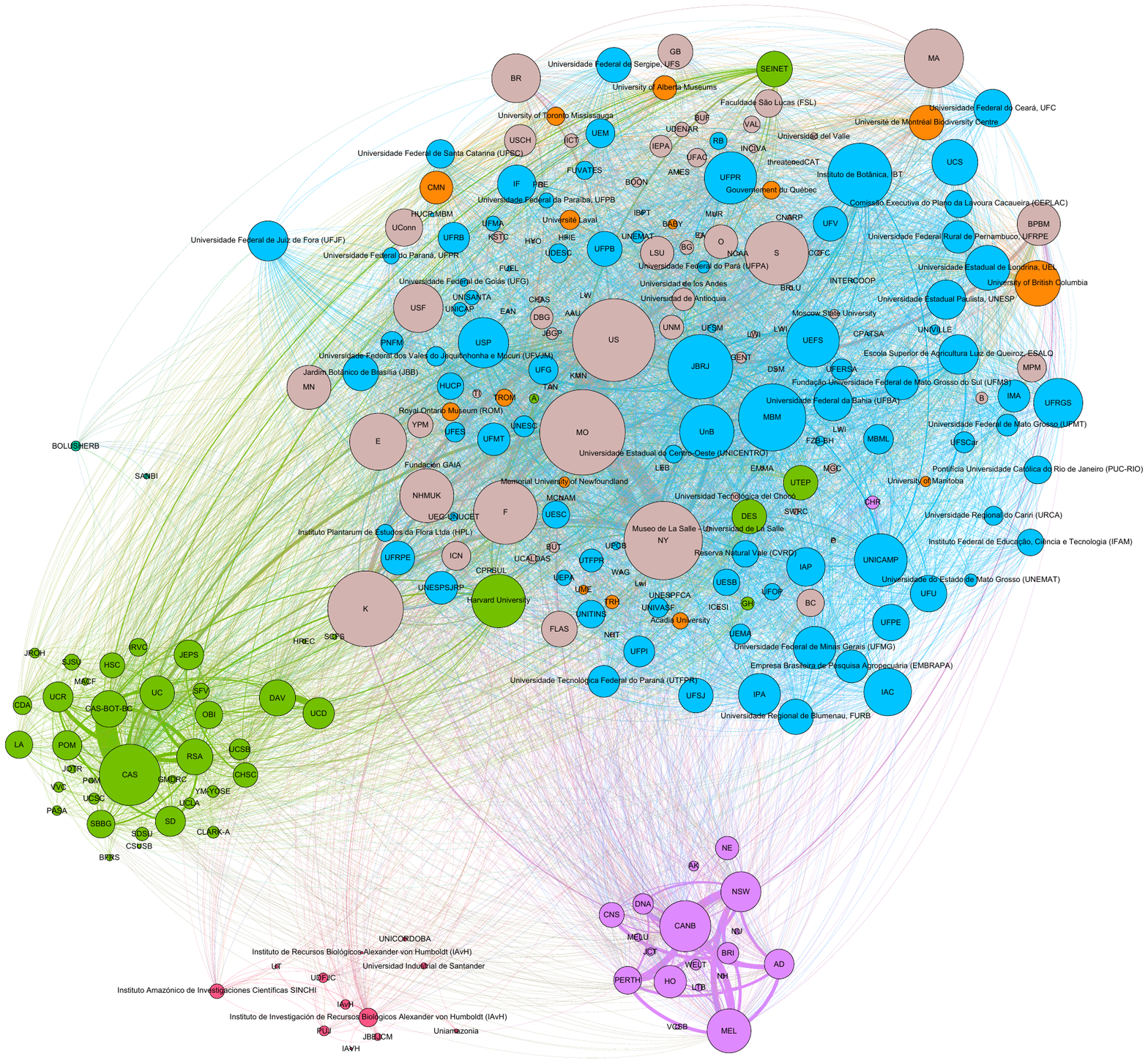}
\caption{Weighted undirected graph from inferred institutional-level
relationships. Nodes represent institutions and are sized by degree (the number
of relationships in which the node participates). Edges (links between nodes)
are shown scaled in proportion to their weight (the number of specimen groups
shared between the two institutions). Seven communities are shown, regional
herbaria: Brazilian (blue), Australasian (purple), United States (green), 
Colombian (red), Canadian (orange) and South African (dark green) with 
the remainder and the international herbaria shown in tan.}\label{fig:inst_graph} 
\end{figure*}

Only the subset of duplicate groups meeting the most conservative
assessment criteria were used in subsequent analyses: 7,102,710
specimens in 2,823,651 groups. The sizes of the conservatively assessed
duplicate groups are shown in figure \ref{fig:dupl_group_sizes}.

\subsection{Propagation of annotations}

Members of duplicate sets are located at different institutional
repositories and therefore may have been curated differently.
Reconciliation of duplicate sets allows the propagation of several
classes of annotations - georeferences, type citations, specimen images
and determinations - between holders. Of the conservatively assessed
duplicate sets:

\begin{itemize}
\tightlist
\item
  93,044 specimens in 54,435 groups could receive a type citation from a
  peer in their duplicate group
\item
  1,121,865 specimens in 782,655 groups could receive a georeference
  from a peer in their duplicate group
\item
  1,097,168 specimens in 758,416 groups could be linked to an associated
  specimen image from a peer in their duplicate group
\item
  2,191,179 specimens are in 792,274 groups which have multiple
  scientific names within the group (indicating uneven scientific name
  determination amongst the members of the specimen duplicate group)
\end{itemize}

\subsection{Repository relationship analysis}

The relationship graph derived from duplication links at institutional
level (see figure \ref{fig:inst_graph}) comprises 260 nodes
(institutions) and 6,588 weighted edges (relationships between
institutions, based on co-participation in a specimen duplicate group,
weighted by the number of co-occurrences). The graph was found to
contain seven communities: Brazilian herbaria, Australasian herbaria,
the regional herbaria in the United States, Colombian herbaria, Canadian
herbaria, South African herbaria, and the internationally focussed
herbaria found in North America and Europe.

\section{Discussion}

\subsection{Duplicate identification and assessment}

A considerable number of duplicate groups were found in the datamined
dataset, and these appear relatively stable across the different
assessment flag combinations (see \ref{fig:assessment_graph}),
permitting the reconciliation of many specimen duplicates between
different specimen repositories. The reconciliation of specimen
duplicate groups show that many metadata annotations could be propagated
between specimen repositories. As these annotations represent both the
most expensive parts of the digitisation process (georeferencing) and
the most valuable kind of usage citation (type citation), mobilising
these between partners would reduce data management costs, improve the
utility of the digitised specimen data and improve institution level
data usage reporting. It is only possible to supply an estimate range
for the cost saving of mobilising such a large number of georeferences.
Standard procedures tend to batch work by locality, which improves
georeferencing speed by focussing on a particular area. A software
description paper reports a project georeferencing at a rate of 16.6 (±
8.3) georeferences per hour and a further separate project achieving a
doubling of this rate \cite{hill_location_2009}. A herbarium type
specimen focussed project reported ``whole process of georeferencing the
ca. 3400 Type specimens took eight months (appx. 100 specimens per
week)'' \cite{garcia2010data}. It seems that there are significant
savings that could be made using the results of this research, given
that the number of propagable georeferences is counted at around a
million.

\subsection{Repository relationship analysis}

The different repositories represented in the dataset are well
connected. Viewed at an institutional level, the low incidence of
isolated cliques shows healthy inter-institutional working relationships
in botany. There are strong links among regionally focussed herbaria in
the United States and Australasia. The interconnections between the
Brazilian herbaria and their international counterparts show the volume
of work that has been focussed on the world's most mega-diverse country
\cite{mittermeier_megadiversity_1997} and also suggest that the data
repatriation projects which aim to mobilise data held out of country
\cite{noauthor_reflora_nodate} have been successful. Quantifying the
links between specimen repositories enables evidence drawn from specimen
duplicate sharing to be used when building project collaborations. Sets
of institutions could be selected to maximise overlap or to maximise
complementarity. Better sharing of specimen data between institutions
facilitates community curation and helps to reduce data management
costs.

\section{Further work}

There are several areas in which future work could develop this analysis
including further refinement of the analytical approach to cover more
data sources, community assessment of interlinked repositories and
quality control of annotations by comparison between duplicates. It may
be useful to separate future work into two streams: a stream regarding
data management and refinement of the data pipeline, and a more
conceptual stream regarding implications of the results. An example from
each area is outlined here: investigation of the reasons why specimens
are not currently identified as duplicates - singleton analysis - and
further work on the research recognition of determination annotations
made on specimen objects.

Singleton specimens may be due to uneven digitisation and / or lack of
participation in data mining process, rather than true singletons,
further data analysis work is required to investigate this. It should be
possible to use the results from the data mining process to calculate
for each collector the likely number of specimens gathered at each
collecting event. These numbers would give us a potential view on the
number of currently un-digitised specimens, and among these, the likely
location of duplicates (in which institutional repositories will they be
found).

Traditional taxonomic activity can be separated into three phases -
collection of specimens, labelling specimen with names and formal
publication of results. The first two phases are absent from traditional
publication focussed career credit, yet generate long-term
research-grade outputs which may be consulted and referenced by others.
As these outputs are now mobilised and used much more widely (due to
data mobilisation via the internet) there have been calls for these to
be included in the career assessment system for taxonomists
\cite{mcdade_biology_2011}. If we recognise that specimens are
persistent research objects, which can be uniformly accessed
\cite{guntsch_actionable_2017}, then the labelling of specimens with
scientific names could each be considered to meet the minimum criteria
for a nanopublication - the smallest unit of research work
\cite{groth_anatomy_2010} and credited to individual researchers.

\section{Conclusion}

Specimens are research objects which are managed for long term
consultation, facilitate scientific discovery and act as vehicles for
the dissemination of results. This paper demonstrates that specimens
form a shared global resource, and that fragmented information
management can be overcome by the reconciliation of specimen duplicates
across institutional boundaries. Specimen digitisation efforts and work
to define standard representations of digitised metadata have built a
critical mass of computable information, which can be used as the input
into this process. Identification of specimen duplicates allows
quantification of potential specimen metadata exchange between
institutional specimen repositories. The result of implementing this
data exchange would be to develop and strengthen ties between
institutional repositories, improve efficiency of data curation (by
eliminating repeated work such as specimen georeferencing) and to
improve the metadata holdings and reporting figures for institutional
repositories. Conceptually, specimens should be recognised as a unit of
research work more granular than the scientific paper, but fulfilling
the same functions - communication of results and establishment of a
long term record. This recognition of the specimen as a research object
would eventually allow the annotation of specimens to be regarded as
research work and credited to individual researchers. This may start to
address some concerns recently voiced with regard to the many phases of
research work conducted by taxonomists which remain absent from
publication-focussed career metrics \cite{mcdade_biology_2011}.

\bibliography{article} 
\bibliographystyle{IEEEtran}


\end{document}